\documentclass[preprint2]{aastex63}

\usepackage{graphicx}
\usepackage[suffix=]{epstopdf}
\usepackage{natbib}
\usepackage{amsmath}
\usepackage{url}
\usepackage{xspace}

\providecommand{\e}[1]{\ensuremath{\times 10^{#1}}}

\begin{document}

\title{
Automated Spectroscopic Wavelength Calibration using 
Dynamic Time Warping 
}

\correspondingauthor{James. R. A. Davenport}
\email{jrad@uw.edu}

\author[0000-0002-0637-835X]{James R. A. Davenport}
\affiliation{Astronomy Department, University of Washington, Box 951580, Seattle, WA 98195, USA}

\author{Francisca Chabour Barra}
\affiliation{Astronomy Department, University of Washington, Box 951580, Seattle, WA 98195, USA}

\author{K. Azalee Bostroem}
\affiliation{Steward Observatory, University of Arizona, 933 North Cherry Avenue, Tucson, AZ 85721-0065, USA}
\altaffiliation{LSST-DA Catalyst Fellow}

\author{Sarah Tuttle}
\affiliation{Astronomy Department, University of Washington, Box 951580, Seattle, WA 98195, USA}

\author{Josue Torres}
\affiliation{Astronomy Department, University of Washington, Box 951580, Seattle, WA 98195, USA}

\author[0000-0002-7961-6881]{Jessica Birky}
\affiliation{Astronomy Department, University of Washington, Box 951580, Seattle, WA 98195, USA}

\author[0000-0003-0484-3331]{Anastasios Tzanidakis}
\affiliation{Astronomy Department, University of Washington, Box 951580, Seattle, WA 98195, USA}

\author{Kal Kadlec}
\affiliation{Astronomy Department, University of Washington, Box 951580, Seattle, WA 98195, USA}

\author[0000-0001-5538-0395]{Yuankun Wang}
\affiliation{Astronomy Department, University of Washington, Box 951580, Seattle, WA 98195, USA}

\author[0000-0002-6629-4182]{Suzanne L. Hawley}
\affiliation{Astronomy Department, University of Washington, Box 951580, Seattle, WA 98195, USA}

\author[0000-0003-3601-3180]{Trevor Z. Dorn-Wallenstein}\thanks{Carnegie Fellow}
\affiliation{Observatories of the Carnegie Institution for Science \\
813 Santa Barbara Street \\
Pasadena, CA 91101, USA}

\author{William Ketzeback}
\affiliation{Apache Point Observatory, 2001 Apache Point Road, Box 59, Sunspot, NM 88349, USA}

\author{Julene Elias}
\affiliation{The College of Idaho, 2112 Cleveland Blvd, Caldwell, ID 83605, USA}

\author{Abdullah Korra}
\affiliation{The College of Idaho, 2112 Cleveland Blvd, Caldwell, ID 83605, USA}

\author{Anna Panova}
\affiliation{The College of Idaho, 2112 Cleveland Blvd, Caldwell, ID 83605, USA}

\author[0000-0002-3723-6362]{Kathryn Devine}
\affiliation{The College of Idaho, 2112 Cleveland Blvd, Caldwell, ID 83605, USA}

\author[0000-0001-6914-7797]{Kevin R. Covey}
\affiliation{Physics \& Astronomy Dept., Western Washington Univ., 516 High St., Bellingham, WA, 98225-9486}

\begin{abstract}
Here we present an automated method for obtaining wavelength calibrations for one-dimensional spectra, using Dynamic Time Warping (DTW). DTW is a flexible and well-understood algorithm for pattern matching, which has not been  widely used in astronomy data analysis. Employing a calibrated template spectrum as a reference, DTW can recover non-linear and even discontinuous dispersion solutions without an initial guess. The algorithm is robust against differing spectral resolution between the template and sample data, and can accommodate some spurious or missing features. We demonstrate the effectiveness of DTW in an automated data reduction workflow, using both simulated and real arc lamp spectra in a Python data reduction framework. Finally, we provide a discussion on the utility and best practices with the DTW algorithm for wavelength calibration. 
We also introduce the PyKOSMOS data reduction toolkit, which includes our DTW calibration methods.
\end{abstract}

\section{Introduction}

Wavelength calibration is a critical and often time consuming task in the standard reduction of astronomical spectroscopy. The traditional approach for wavelength calibration is to use exposures of arc lamps as sources with known wavelength features (i.e. narrow emission lines). Individual emission lines are typically identified by eye, often while comparing to a reference arc lamp spectrum or a library of strong lines. 
By using arc lamps with many prominent emission lines across the full range of wavelength sampled by the spectrograph, users can solve for even non-linear dispersion solutions. The final wavelength solution (i.e. transforming instrumental pixel coordinates into wavelength) is produced by interpolating between the identified emission lines. 

The interactive line identification process has long been a challenge for observers, as it is tedious and error prone. Misidentifying a line, or identifying too few lines, can result in poor wavelength solutions. Often many iterations of line identification and model fitting are often needed to produce a sufficiently accurate wavelength solution, even for well tested spectrographs. Traditionally this has been carried out using the {\tt identify} package within {\tt IRAF}. Hands-free wavelength calibration methods have been developed, which range from updating solutions assuming small perturbations ({\tt reidentify} in {\tt IRAF}), to fully automated solutions, and using a variety of pattern matching algorithms \citep[e.g.][]{valdes1996}. Similarly, large spectroscopic surveys have developed highly tuned, automated methods for producing accurate and precise wavelength solutions every night \citep[e.g.][]{stoughton2002}. Still, manual line identification and wavelength solution modeling remains a common part of the data reduction workflow.

Dynamic Time Warping (DTW) is a flexible pattern matching algorithm with a long history of use in many domains. DTW was popularized initially in speech recognition, proving effective at matching waveforms while accounting for varying rates or volumes while speaking \citep{velichko1970}, and has been used in countless other pattern matching applications. Though it has seen limited use in astronomy so far, DTW has for example been used in aligning time series measurements of the solar wind from multiple sensors \citep{samara2022}. DTW has also been used for calibrating laboratory spectroscopy in the near infrared in the field of chemistry \citep{zou2019}.

Here we introduce DTW as a straightforward means for obtaining wavelength calibrations for optical spectroscopy. The flexibility of its pattern matching means that, given an appropriate template, DTW can solve fairly complex dispersion solutions without any interactive feature identification. With a modest development of templates, and using existing implementations of the algorithm, we believe DTW can provide a general solution for automated wavelength calibration in most standard data reduction purposes.


\section{Dynamic Time Warping}
\label{sec:DTW}

The Dynamic Time Warping (DTW) class of algorithms are used in many domains to align datasets, or match data to a library of models, by stretching or compressing (i.e. warping) them.
A more detailed description of the DTW algorithm can be found in \citet{senin2008}. Various optimization and performance improvements are described in \citet{DTW}. 
Applications for DTW in matching patterns within time series data is given in \citet{berndt1994}, and for hand writing or signature recognition by \citet{munich1999}. Here we briefly summarize relevant properties of the DTW method.

\begin{figure}[]
\centering
\includegraphics[width=3.5in]{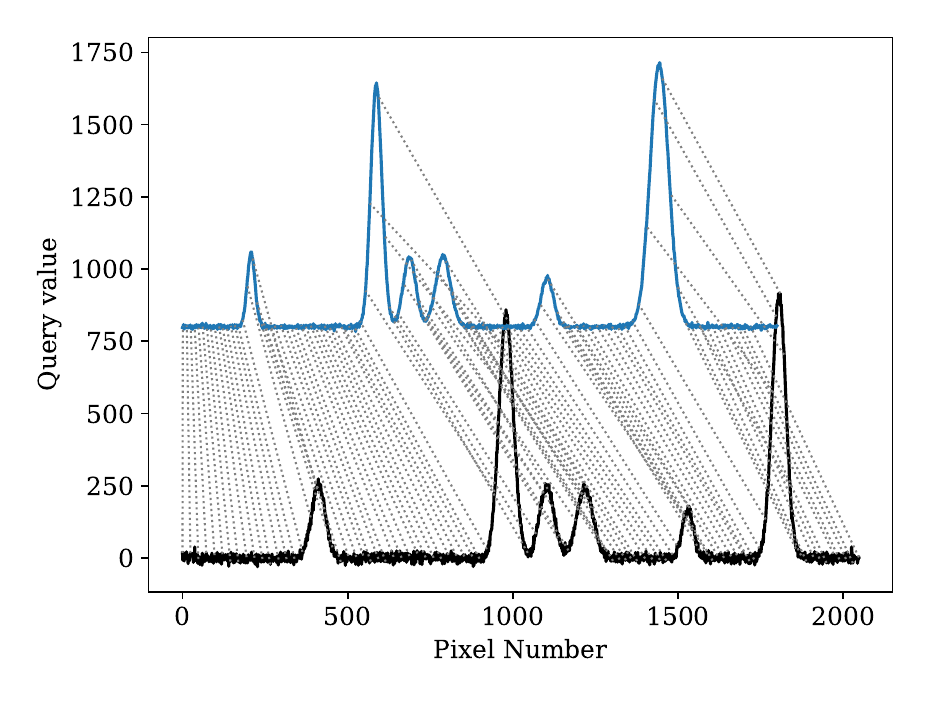}
\caption{
Example of a DTW alignment for a simulated ``reference'' spectrum (blue) and a ``query'' spectrum (black) that has undergone a logarithmic expansion of the wavelength axis. Dotted lines illustrate the pixel-to-pixel alignment computed using DTW.
}
\label{fig:cartoon}
\end{figure}

As illustrated in Figure \ref{fig:cartoon}, the amount of warping is determined locally. DTW aligns individual features in the data (e.g. an emission line in a spectrum). For example, in the classical use of DTW for speech recognition, the algorithm can reliably match reference recordings for the word ``{\it space}'' with a given sample word ``{\it spaaaaace}'', despite having an exaggerated `a' sound.
DTW ultimately produces a distance metric between the observed data and the reference, which quantifies how similar they are after the alignment is found. In typical model-fitting fashion, the reference with the smallest distance to the data is considered the ``best fit''. 

Figure \ref{fig:cartoon} demonstrates the classical DTW solution where the entire ``reference'' (i.e. template) is matched to the entire ``query'' (i.e. observation). Methods for open-ended alignment are also available, which allow matching of the data to a subsection of a larger template. In this paper we use the DTW toolkit available for R and Python developed by \citet{DTW}, which provides a wide variety of alignment algorithms and distance metrics, including methods that can align reversed or highly distorted data.

Since it is fairly robust against noise, non-linear alignments, and differing data resolution, DTW is well suited for aligning spectra to models or templates. DTW succeeds where many standard model alignment approaches such as cross-correlation fail, with warped or non-linear stretches. As we demonstrate here, DTW can easily be used to identify spectral lines, and solving for wavelength calibrations in data.

\section{Applying DTW to Spectral Calibration}
\label{sec:cal}

One common application where relatively simple pattern matching is critical to traditional data analysis workflows is in the wavelength calibration of spectroscopy using emission line lamps. The manual identification of emission or arc lines to develop a dispersion solution has been a part of standard spectroscopic reduction for decades, and is a well developed part of the IRAF workflow \citep{tody1986}. However, this is time consuming, and prone to human errors in data entry and precise line identification from reference images or tables.

Automating this wavelength calibration step has been implemented in several ways. In IRAF, the \texttt{REIDENTIFY} method allows previous wavelength calibrations to be shifted by a few pixels to automatically account for small amplitude (few pixel) shifts over the course of a night. The \texttt{AUTOIDENTIFY} method uses a pattern matching algorithm based on the ratios of emission line separations in wavelength, which should be robust against most linear dispersion solutions \citep{valdes1996}. Industrial spectroscopic surveys such as the SDSS MaNGA program that obtain hundreds to thousands of spectra simultaneously use highly tuned methods, such as cross-correlating an arc lamp spectrum against a purpose built model \citep{law2016}. With such an automated procedure, MaNGA is able to efficiently calibrate every science exposure, and produce wavelength calibrations accurate to $\sim$5 km s$^{-1}$.

A DTW-based wavelength calibration is achieved in much the same way as these existing techniques. Like the methods listed above, DTW requires a reference spectrum or model with a reliable wavelength solution that can be aligned to the user's data. The query spectrum (i.e. data to calibrate) must of course be contained within the wavelength domain of the reference spectrum, and they must share similar emission line features. While DTW can align data with missing or exaggerated features, differing signal-to-noise, or data resolutions compared to the reference, attempting to align data from a very different model (e.g. using the wrong emission line template) will give spurious results. Since DTW uses the entire spectrum for alignment, even small amplitude features (e.g. weak lines) can help in the solution if the template has comparable resolution and instrumental broadening.

Our two-step approach to solving for wavelength dispersion using DTW works as follows: 
We first obtain emission line spectra for our data (the ``query'') and for an appropriate template (the ``reference''), using standard boxcar extraction. Each spectrum is normalised by the spectrum's median flux to give comparable emission line amplitudes regardless of exposure time. The fluxes from these two spectra are passed to the Python implementation of the DTW alignment code from \citet{DTW}. 
We allow the query spectrum to be aligned to a subsection of the reference spectrum using the \texttt{open\_begin} and \texttt{open\_end} keywords, and using setting the alignment model option \texttt{step\_pattern=`asymmetric'}. The DTW algorithm then returns the best alignment that maps the query to the reference spectrum.

Since DTW uses the entire spectrum in its alignment, it can produce discontinuous or ``jumpy'' solutions, particularly for data with low signal-to-noise. We note this can result in non-physical wavelength solutions between the emission lines, where there is little information for DTW to align to. Some DTW solving algorithms allow priors or constraints to be placed on the warping, for example limiting the maximum distance to stretch, or penalizing solutions with rapid changes in the alignment. Tuning these constraints is specific to the use case, and beyond the scope of this paper to explore. 
For arc spectra with narrow emission lines such as we focus on here, we do not recommend using DTW alone without either constraining the alignment, or doing a second-pass analysis on the resulting alignment as we demonstrate.

We focus on the emission lines in the second step of our DTW-based calibration algorithm, since they represent the most information-rich portions of the arc lamp spectra. 
We use the \texttt{find\_peaks} method in \texttt{scipy} to find all strong peaks, requiring that peaks are at least 5 pixels apart. Since this function only identifies the highest pixel, we centroid each emission line using a Gaussian profile using a standard least squares fitting. 
This selects primarily strong emission lines, though they can be much lower in amplitude than the strongest lines often identified in manual analysis. We assume that the DTW alignment is successful for these lines, and that they have reliable wavelengths mapped from the reference to the query spectrum. We finally fit a smooth model to the aligned wavelengths for these emission lines, which can include a polynomial, spline, or Gaussian Process. This is exactly akin to the final step in the IRAF \texttt{IDENTIFY} procedure, and provides the wavelength dispersion solution used to calibrate our data.

\section{Solving Complex Dispersions}
\label{sec:model}

\begin{figure}[]
\centering
\includegraphics[width=3.5in]{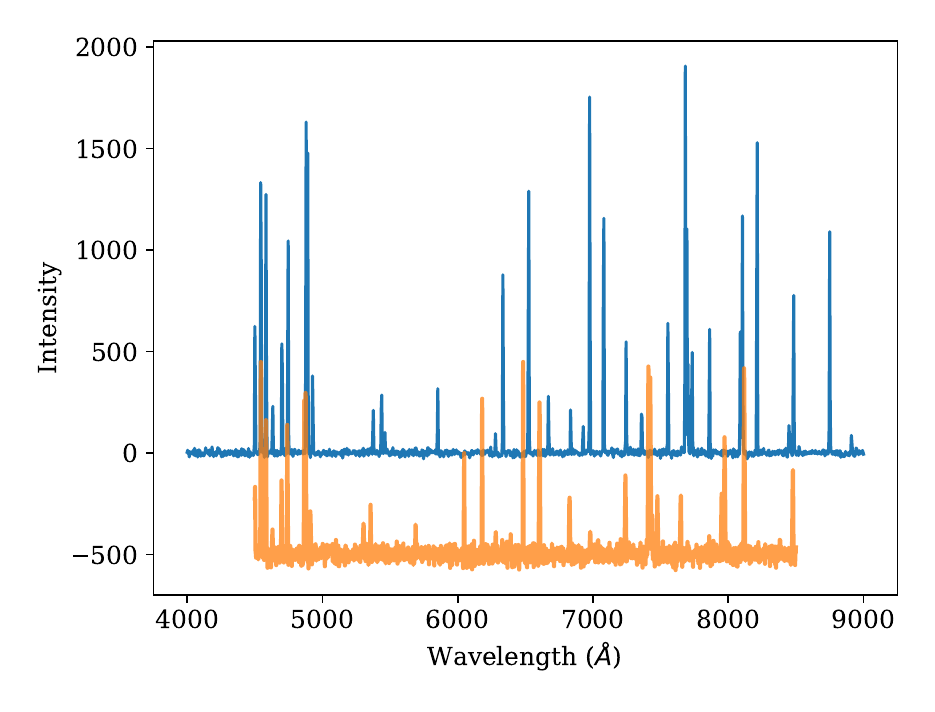}
\caption{Synthetic arc lamp with 40 random emission lines and a small amount of Gaussian noise (blue line). A mock observation with lower resolution, lower signal-to-noise, 10 missing emission lines, and a discontinuous wavelength stretch is shown below (orange line). Note the mock observation has been arbitrarily aligned in wavelength and no wavelength solution has been computed here. 
}
\label{fig:model1}
\end{figure}

To demonstrate the utility of our DTW-based algorithm for automated wavelength calibration, we have generated a dispersion scenario that should be much more difficult than is common.
We assume a spectrum with a rich set of emission lines, as would be expected for many common arc lamps (e.g. Neon), but do not explore the limits of DTW for matching arbitrary patterns with small numbers of features. For this simulation we created a synthetic arc lamp reference spectrum, using 30 emission lines with random amplitudes and wavelengths between 4000 and $9000$\AA, and assuming a Gaussian profile width of 2\AA for each emission line. This toy model was developed to represent a typical arc lamp observed in the optical regime with a low resolution ($R\approx1200$) instrument. 
The wavelength scale of the reference spectrum was linear, with 1280 pixels.
A small amount of Gaussian white noise was finally added to the reference flux, which is shown in Figure \ref{fig:model1}.

We then created a mock observation of this synthetic emission lamp. This was done by first interpolating the reference flux down to 1024 pixels between 4500--8500\AA, assuming a piece-wise wavelength dispersion with logarithmic scaling up to 7000\AA, and linear scaling up to 8500\AA. While a discontinuous wavelength solution is not a common situation for most spectrographs, this example serves to demonstrate the power of DTW to perform in situations that simple cross correlation cannot handle.
White noise was added with 2.5 times larger amplitude than the reference spectrum, simulating a lower signal-to-noise from a routine arc lamp spectrum compared to a high fidelity template.
Finally, to further confound the solution, 10 additional lines were placed into the {\it reference} spectrum. This simulates a reference template with two elements (e.g. Neon and Krypton), but with  an observation of only Neon. The observed spectrum is also shown in Figure \ref{fig:model1}, with an arbitrary alignment for comparison (i.e. no wavelength solution is computed). While some major clusters of emission lines can be identified by eye, this is a very difficult spectrum to manually identify emission lines from and produce a reliable dispersion solution.

\begin{figure}[]
\centering
\includegraphics[width=3.5in]{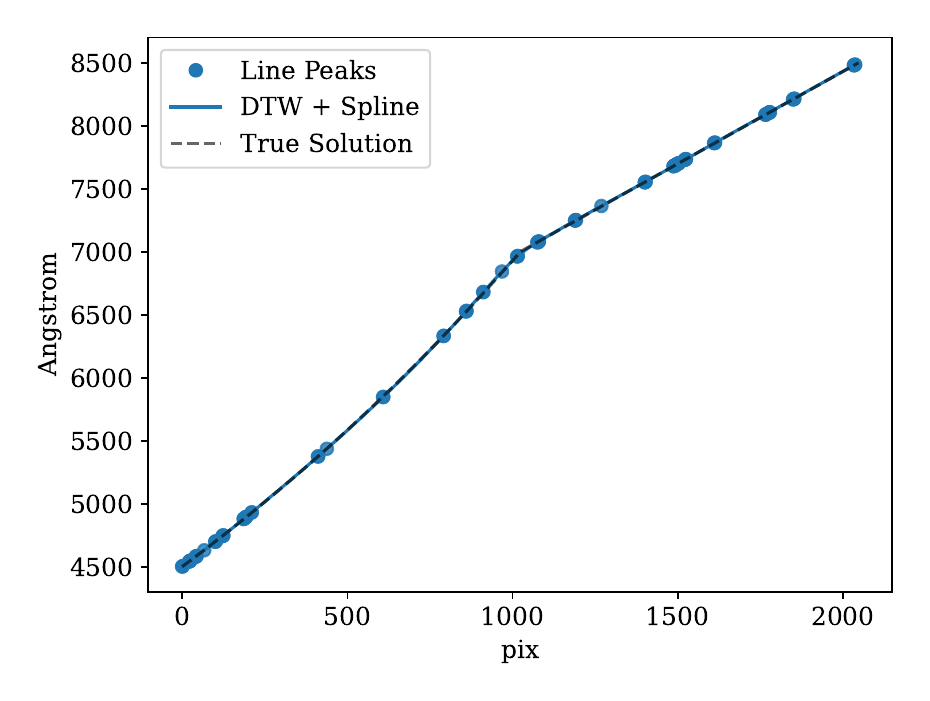}
\caption{Dispersion solution from our two-step DTW algorithm for the simulated arc lamp shown in Figure \ref{fig:model1}. This example was intentionally difficult, with a discontinuity at 7000\AA. The query spectrum was missing 25\% of the emission lines in the reference spectrum, and had lower signal to noise. Emission line peaks are robustly identified (blue dots). A spline model is fit to produce the final dispersion solution (blue solid line), which closely matches the input simulation (black dashed line).
}
\label{fig:modelfit}
\end{figure}

In Figure \ref{fig:modelfit} we show the result of our two-step DTW fitting procedure described in \S \ref{sec:cal}. As expected, the local alignment from DTW provides an adaptive solution, with a clear discontinuity around 7000\AA\ as was input into the mock observations. Prominent emission lines have then been identified, and a cubic spline model has been fit to produce the final solution. This challenging scenario was computed in 1.6 seconds on a typical laptop computer, with no interactivity or fine tuning beyond the standard peak detection threshold of 95\%. Since the emission lines were randomly generated in amplitude and wavelength, the ability to precisely model the discontinuity point is dependent on the density of strong lines near the break wavelength of {7000\AA}. 
As is well known, spline fits to dispersion solutions can also become unstable at the edges of the spectrum, and the user may prefer a polynomial or a Gaussian Process fit.

\section{Demonstrating with Actual Data}
\label{sec:kosmos}
\subsection{Fitting KOSMOS Data}

\begin{figure*}[]
\centering
\includegraphics[width=3.5in]{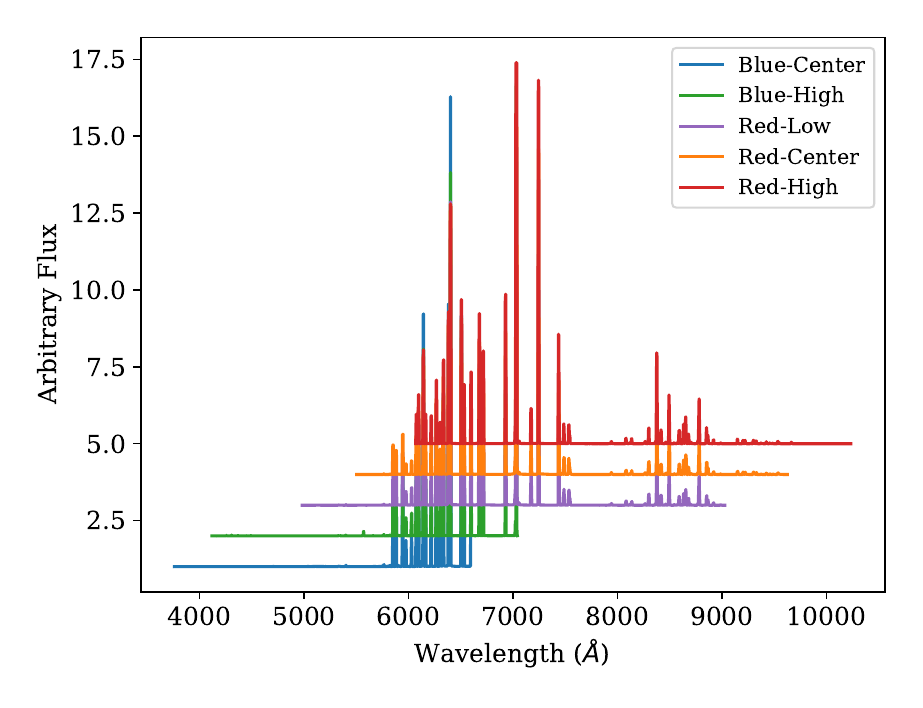}
\includegraphics[width=3.5in]{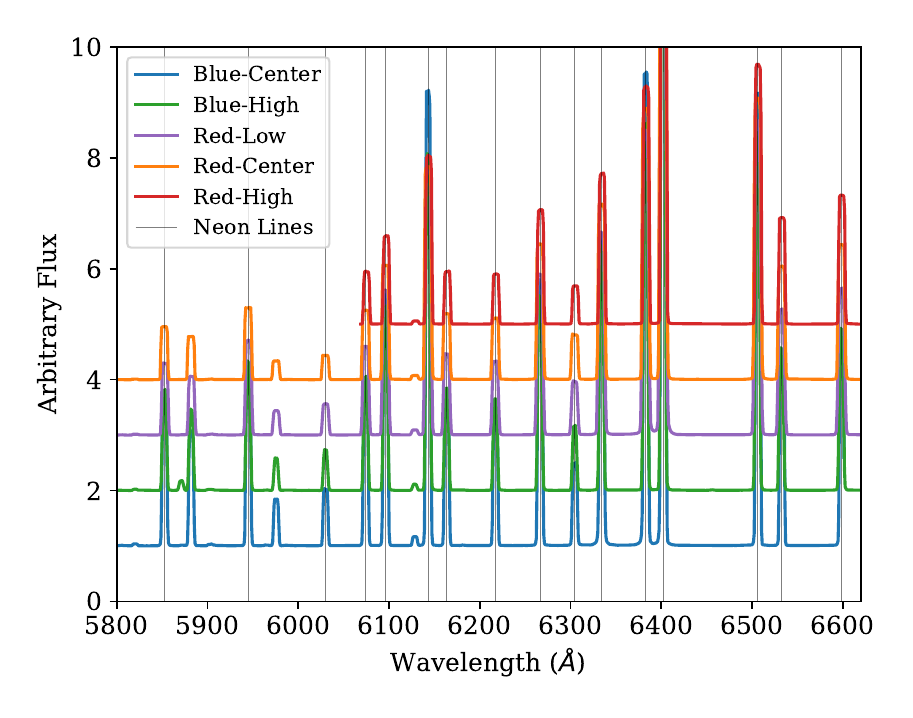}
\caption{
Left: Neon spectra from the internal arc lamp in five of the available grism modes from KOSMOS on APO. These data were automatically aligned with our two-step DTW algorithm, using the manually calibrated templates we developed for KOSMOS.
Right: Zoomed in portion of the same aligned spectra, with known Neon emission lines from IRAF's {\tt idhenear} table overlaid for reference. Our DTW algorithm successfully transfers the dispersion solution from the reference template to all five of the query spectra shown.
}
\label{fig:soln1}
\end{figure*}

Beyond the performance of DTW using simulated emission lamps, we have done extensive testing of our algorithm using data from the KOSMOS longslit spectrograph on the Apache Point Observatory (APO) 3.5-m telescope \citep{tran2019}. KOSMOS has multiple built-in arc lamps for easy wavelength calibration, which presently include a Neon, Argon, and Krypton lamp that can be illuminated in any combination. We obtained arc spectra for all three of these lamps using the six available standard grism modes for KOSMOS, using 1x1 pixel binning with a total of 4096 pixels along the wavelength axis. All prominent emission lines were manually identified following  standard IRAF \texttt{IDENTIFY} procedures, and template arc spectra for KOSMOS were then saved. 

These arc templates are ideal for calibrating KOSMOS data using our DTW-based algorithm.
In Figure \ref{fig:soln1} we demonstrate the successful alignment of Neon lamps for the five reddest grism modes based on these templates and our two-step DTW-based alignment algorithm. The ``Blue Low'' grism mode did not produce a robust wavelength solution since there were insufficient Neon lines present. The known Neon lines from the IRAF {\tt idhenear} catalog  match these query spectra, despite being only used in the alignment of the reference spectra. This demonstrates the ability to robustly determine the wavelength solution via DTW with templates derived from the same instrument.

\subsection{Applying KOSMOS Templates to DIS}
\label{sec:dishi}

Though the spectral templates above are best used with data from the same instrument they were generated with, they can be used to successfully align data from different spectrographs, or equivalently from the same spectrograph but in a different configuration. To demonstrate this capability, we use a separate spectrograph that is located on the APO 3.5-m.
The Double Imaging Spectrograph (DIS) is a well tested instrument that was a pathfinder for the Sloan Digital Sky Survey's original spectrograph \citep{york2000}. This venerable instrument is the ideal test for aligning KOSMOS templates using DTW.

\begin{figure}[]
\centering
\includegraphics[width=3.5in]{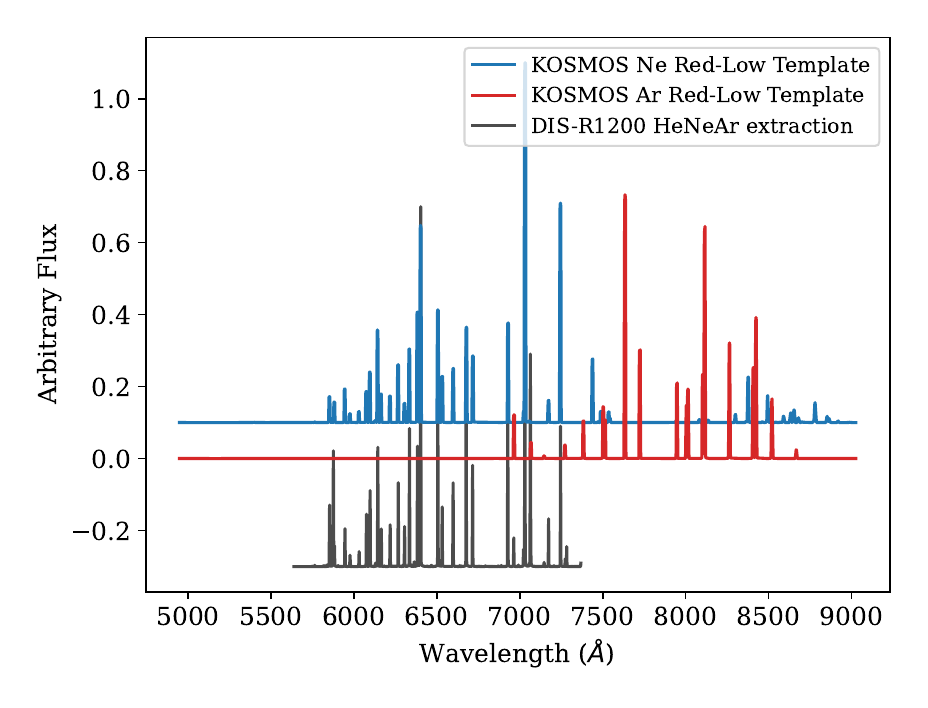}
\caption{
HeNeAr lamp spectrum from DIS in the ``R1200'' medium-resolution mode (black line), aligned with our DTW-based algorithm using a combination of KOSMOS Neon (blue curve) and Argon templates (red curve). Despite missing a small number of Helium features, and different instrumental properties, the KOSMOS templates can generate a reliable wavelength calibration.
}
\label{fig:dishigh}
\end{figure}

In Figure \ref{fig:dishigh} we show an example of a standard combined Helium + Neon + Argon calibration lamp spectrum using the ``R1200'' medium resolution mode ($R\sim7000$) with DIS. Since KOSMOS at present does not contain an internal Helium arc lamp, we used templates for Neon and Argon. The ``Red-Low'' KOSMOS templates were used, since they covered the wavelength range expected for the DIS R1200. 
Despite missing the Helium lines, and having different resolutions and instrument systematics, our DTW-based algorithm successfully aligns the DIS data to the KOSMOS template.
We note that the KOSMOS Neon and Argon templates here were simply codded, and we did not try to scale the Neon and Argon KOSMOS line amplitudes to match the DIS spectrum. However, if the relative emission line amplitudes were drastically different between DIS and KOSMOS, such re-normalization may be needed.

\subsection{Template Limitations}
While our DTW-based algorithm is capable of aligning spectra and templates derived from different instruments, or at various resolutions, it can fail when the instrumental properties are very different. Here we demonstrate an example where our DTW-based algorithm fails to produce a reliable alignment, using the same comparison instrument as before but in a lower resolution mode. We used DIS in the low-resolution ``R300'' mode, observing the Helium + Neon + Argon arc lamps simultaneously. We attempted to align this spectrum with the the same combined Neon and Argon KOSMOS template as in \S \ref{sec:dishi}, which should cover approximately the entire wavelength range of DIS R300. 

\begin{figure}[]
\centering
\includegraphics[width=3.5in]{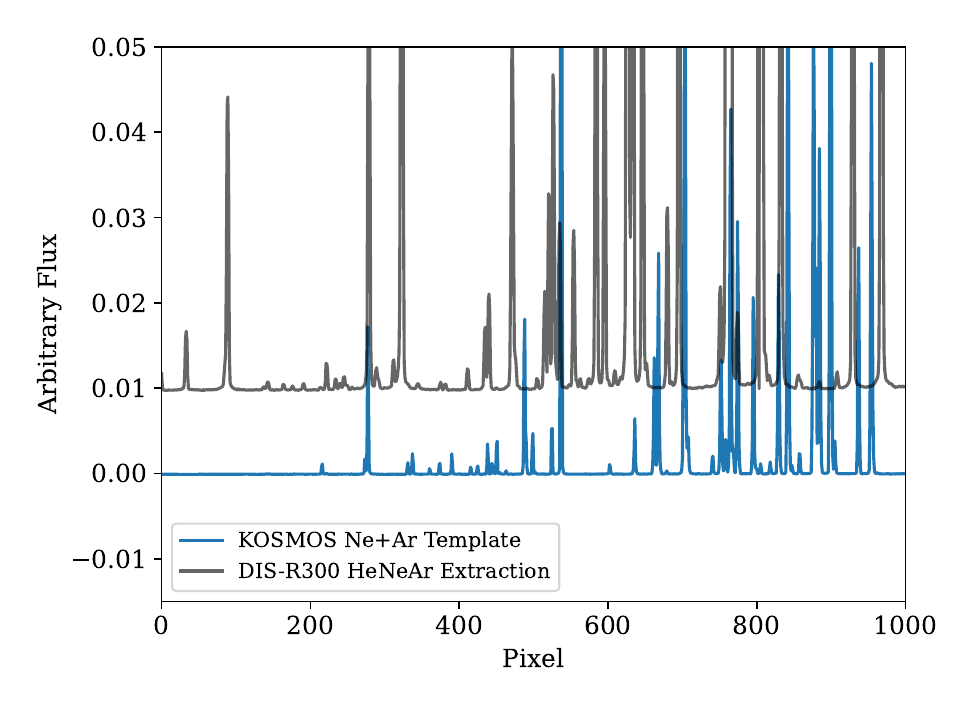}
\caption{
A portion of the HeNeAr lamp spectrum from DIS in the ``R300'' low resolution mode (black line). Even when down-sampling the templates to approximately match the resolution of the query spectrum as shown, KOSMOS Neon + Argon templates fail to produce a reliable dispersion solution using our algorithm. Since DTW uses the entire spectrum, not only the locations of the peaks, it fails to accurately align these curves due to the significantly different emission line amplitudes and profiles as shown. The DIS R300 spectrum can be reliably aligned with our DTW algorithm using a more appropriate template.
}
\label{fig:dislow}
\end{figure}

As shown in Figure \ref{fig:dislow}, our DTW-based algorithm is unable to correctly align DIS R300 spectrum using the KOSMOS-derived template. This incorrect alignment arises because DTW finds the best match for the entire sample spectrum (here the DIS R300), without regard for specific features (e.g. emission lines). The DIS-R300 spectrum is visibly dominated by extended broadening due to scattered light in the instrument, and the lower spectral resolution. It is possible to algin the DIS-R300 data with the KOSMOS template by down-sampling both spectra to minimize the impact of line broadening. It may be possible to develop spectral templates and down-sampling or filtering schemes that are very robust against these instrumental systematics. This is beyond the scope of our current work, however, and we recommend generating spectral templates for individual instruments.

\section{Conclusions}
\label{sec:conclusions}

We have presented a simple algorithm for producing wavelength solutions for low- to medium-resolution optical spectroscopy, using Dynamic Time Warping (DTW) to align template arc lamp spectra. This approach is fast, hands free, and reliable in most cases. DTW can solve for non-linear or discontinuous dispersion solutions, and handle some missing features (emission lines). If appropriate templates are used, i.e. generated for the same instrument as the data, we believe our DTW-based algorithm can be reliably used in most data reduction scenarios. We recommend observatories provide calibrated arc lamp spectra in common instrumental setups.

Our algorithm is a hybrid approach, using DTW to develop the first-pass alignment, and then a more traditional model fit to strong emission lines. This avoids problems with a pixel-level stretch in DTW, that can lead to non-physical solutions in the first-pass. An alternative approach could be to use a continuous dynamic time warping algorithm, which is constrained to give more physical solutions \citep[e.g.][]{munich1999}.

Our goal has been to develop a fast and easy to use wavelength solution algorithm to replace the traditional IRAF \texttt{IDENTIFY} function. We have not attempted to develop the highest fidelity wavelength solution possible, such as those created for ``extreme precision radial velocity'' (EPRV) projects \citep[e.g.][]{petersburg2020}. These EPRV instruments are usually high-order, cross-dispersed echelle spectrographs. DTW could be used to align each extracted order of an echelle spectrum, given an appropriate template.

While aligning arc lamp spectra is a straightforward use case for DTW, we emphasize the algorithm has not been tuned for this work. Template matching or alignment algorithms like DTW have a variety of possible new applications within astronomy. These might include quickly identifying spectra types for stars without the need for precise flux calibration \citep{bochanski_templates,roulston2020}, estimating redshifts for galaxies \citep{bolton2012}, or classifying variable star light curves \citep{jayasinghe2018a}. DTW-based approaches could therefore be a valuable, simple addition to the growing suite of machine learning methods in use in astronomy pipelines.

\acknowledgments

JRAD acknowledges support from the DiRAC Institute in the Department of Astronomy at the University of Washington. The DiRAC Institute is supported through generous gifts from the Charles and Lisa Simonyi Fund for Arts and Sciences, Janet and Lloyd Frink, and the Washington Research Foundation.

Based on observations obtained with the Apache Point Observatory 3.5-meter telescope, which is owned and operated by the Astrophysical Research Consortium.

KAB is supported through the LSST-DA Catalyst Fellowship project; this publication was thus made possible through the support of Grant 62192 from the John Templeton Foundation to LSST-DA.

\software{
Python, IPython \citep{ipython}, 
NumPy \citep{numpy}, 
Matplotlib \citep{matplotlib}, 
SciPy \citep{scipy}, 
Pandas \citep{pandas}, 
Astropy \citep{astropy,astropy:2022},
CCDproc \citep{ccdproc},
specutils \citep{specutils},
specreduce \citep{specreduce},
george \citep{george-gp},
DTW \citep{DTW},
PyKOSMOS \citep{pykosmos}
}

\bibliography{references.bib}

\begin{thebibliography}{}
\expandafter\ifx\csname natexlab\endcsname\relax\def\natexlab#1{#1}\fi
\providecommand{\url}[1]{\href{#1}{#1}}
\providecommand{\dodoi}[1]{doi:~\href{http://doi.org/#1}{\nolinkurl{#1}}}
\providecommand{\doeprint}[1]{\href{http://ascl.net/#1}{\nolinkurl{http://ascl.net/#1}}}
\providecommand{\doarXiv}[1]{\href{https://arxiv.org/abs/#1}{\nolinkurl{https://arxiv.org/abs/#1}}}

\bibitem[{{Ambikasaran} {et~al.}(2015){Ambikasaran}, {Foreman-Mackey},
  {Greengard}, {Hogg}, \& {O'Neil}}]{george-gp}
{Ambikasaran}, S., {Foreman-Mackey}, D., {Greengard}, L., {Hogg}, D.~W., \&
  {O'Neil}, M. 2015, IEEE Transactions on Pattern Analysis and Machine
  Intelligence, 38, 252, \dodoi{10.1109/TPAMI.2015.2448083}

\bibitem[{{Astropy Collaboration} {et~al.}(2013){Astropy Collaboration},
  {Robitaille}, {Tollerud}, {Greenfield}, {Droettboom}, {Bray}, {Aldcroft},
  {Davis}, {Ginsburg}, {Price-Whelan}, {Kerzendorf}, {Conley}, {Crighton},
  {Barbary}, {Muna}, {Ferguson}, {Grollier}, {Parikh}, {Nair}, {Unther},
  {Deil}, {Woillez}, {Conseil}, {Kramer}, {Turner}, {Singer}, {Fox}, {Weaver},
  {Zabalza}, {Edwards}, {Azalee Bostroem}, {Burke}, {Casey}, {Crawford},
  {Dencheva}, {Ely}, {Jenness}, {Labrie}, {Lim}, {Pierfederici}, {Pontzen},
  {Ptak}, {Refsdal}, {Servillat}, \& {Streicher}}]{astropy}
{Astropy Collaboration}, {Robitaille}, T.~P., {Tollerud}, E.~J., {et~al.} 2013,
  \aap, 558, A33, \dodoi{10.1051/0004-6361/201322068}

\bibitem[{{Astropy Collaboration} {et~al.}(2022){Astropy Collaboration},
  {Price-Whelan}, {Lim}, {Earl}, {Starkman}, {Bradley}, {Shupe}, {Patil},
  {Corrales}, {Brasseur}, {N{"o}the}, {Donath}, {Tollerud}, {Morris},
  {Ginsburg}, {Vaher}, {Weaver}, {Tocknell}, {Jamieson}, {van Kerkwijk},
  {Robitaille}, {Merry}, {Bachetti}, {G{"u}nther}, {Aldcroft},
  {Alvarado-Montes}, {Archibald}, {B{'o}di}, {Bapat}, {Barentsen}, {Baz{'a}n},
  {Biswas}, {Boquien}, {Burke}, {Cara}, {Cara}, {Conroy}, {Conseil}, {Craig},
  {Cross}, {Cruz}, {D'Eugenio}, {Dencheva}, {Devillepoix}, {Dietrich},
  {Eigenbrot}, {Erben}, {Ferreira}, {Foreman-Mackey}, {Fox}, {Freij}, {Garg},
  {Geda}, {Glattly}, {Gondhalekar}, {Gordon}, {Grant}, {Greenfield}, {Groener},
  {Guest}, {Gurovich}, {Handberg}, {Hart}, {Hatfield-Dodds}, {Homeier},
  {Hosseinzadeh}, {Jenness}, {Jones}, {Joseph}, {Kalmbach}, {Karamehmetoglu},
  {Ka{l}uszy{'n}ski}, {Kelley}, {Kern}, {Kerzendorf}, {Koch}, {Kulumani},
  {Lee}, {Ly}, {Ma}, {MacBride}, {Maljaars}, {Muna}, {Murphy}, {Norman},
  {O'Steen}, {Oman}, {Pacifici}, {Pascual}, {Pascual-Granado}, {Patil},
  {Perren}, {Pickering}, {Rastogi}, {Roulston}, {Ryan}, {Rykoff}, {Sabater},
  {Sakurikar}, {Salgado}, {Sanghi}, {Saunders}, {Savchenko}, {Schwardt},
  {Seifert-Eckert}, {Shih}, {Jain}, {Shukla}, {Sick}, {Simpson},
  {Singanamalla}, {Singer}, {Singhal}, {Sinha}, {Sip{H{o}}cz}, {Spitler},
  {Stansby}, {Streicher}, {{{S}}umak}, {Swinbank}, {Taranu}, {Tewary},
  {Tremblay}, {Val-Borro}, {Van Kooten}, {Vasovi{'c}}, {Verma}, {de Miranda
  Cardoso}, {Williams}, {Wilson}, {Winkel}, {Wood-Vasey}, {Xue}, {Yoachim},
  {Zhang}, {Zonca}, \& {Astropy Project Contributors}}]{astropy:2022}
{Astropy Collaboration}, {Price-Whelan}, A.~M., {Lim}, P.~L., {et~al.} 2022,
  apj, 935, 167, \dodoi{10.3847/1538-4357/ac7c74}

\bibitem[{Berndt \& Clifford(1994)}]{berndt1994}
Berndt, D.~J., \& Clifford, J. 1994, in Proceedings of the 3rd International
  Conference on Knowledge Discovery and Data Mining, AAAIWS'94 (AAAI Press),
  359--370

\bibitem[{{Bochanski} {et~al.}(2007){Bochanski}, {West}, {Hawley}, \&
  {Covey}}]{bochanski_templates}
{Bochanski}, J.~J., {West}, A.~A., {Hawley}, S.~L., \& {Covey}, K.~R. 2007,
  \aj, 133, 531, \dodoi{10.1086/510240}

\bibitem[{{Bolton} {et~al.}(2012){Bolton}, {Schlegel}, {Aubourg}, {Bailey},
  {Bhardwaj}, {Brownstein}, {Burles}, {Chen}, {Dawson}, {Eisenstein}, {Gunn},
  {Knapp}, {Loomis}, {Lupton}, {Maraston}, {Muna}, {Myers}, {Olmstead},
  {Padmanabhan}, {P{\^a}ris}, {Percival}, {Petitjean}, {Rockosi}, {Ross},
  {Schneider}, {Shu}, {Strauss}, {Thomas}, {Tremonti}, {Wake}, {Weaver}, \&
  {Wood-Vasey}}]{bolton2012}
{Bolton}, A.~S., {Schlegel}, D.~J., {Aubourg}, {\'E}., {et~al.} 2012, \aj, 144,
  144, \dodoi{10.1088/0004-6256/144/5/144}

\bibitem[{Craig {et~al.}(2023)Craig, Crawford, Seifert, Robitaille, Sip{\H
  o}cz, Walawender, Crawford, Vin{\'\i}cius, Ninan, Droettboom, Bowers, Youn,
  Gondhalekar, Tollerud, Lim, Bray, Bach, stottsco, Janga, walkerna22,
  G{\"u}nther, Rol, A., Bradley, Price-Whelan, Deil, Ryon, Lee, Barbary, \&
  Weiner}]{ccdproc}
Craig, M., Crawford, S., Seifert, M., {et~al.} 2023, astropy/ccdproc: 2.4.1,
  2.4.1,  Zenodo, \dodoi{10.5281/zenodo.7986923}

\bibitem[{Davenport \& Bostroem(2023)}]{pykosmos}
Davenport, J., \& Bostroem, A. 2023, {jradavenport/pykosmos: PyKOSMOS: An easy
  to use reduction package for one-dimensional longslit spectroscopy.}, v0.3.1,
   Zenodo, \dodoi{10.5281/zenodo.10152906}

\bibitem[{Earl {et~al.}(2023)Earl, Tollerud, O'Steen, brechmos, Kerzendorf,
  Busko, shaileshahuja, D'Avella, Robitaille, Lim, Ginsburg, Homeier, Sip{\H
  o}cz, Averbukh, Tocknell, Cherinka, Ogaz, Geda, Davies, Conroy, G{\"u}nther,
  Barbary, Foster, Droettboom, Nguyen, Bray, Casey, Teuben, Crawford, \&
  Ferguson}]{specutils}
Earl, N., Tollerud, E., O'Steen, R., {et~al.} 2023, astropy/specutils: v1.12.0,
  v1.12.0,  Zenodo, \dodoi{10.5281/zenodo.10016569}

\bibitem[{Giorgino(2009)}]{DTW}
Giorgino, T. 2009, Journal of Statistical Software, Articles, 31, 1,
  \dodoi{10.18637/jss.v031.i07}

\bibitem[{Hunter(2007)}]{matplotlib}
Hunter, J.~D. 2007, Computing In Science \& Engineering, 9, 90,
  \dodoi{10.1109/MCSE.2007.55}

\bibitem[{{Jayasinghe} {et~al.}(2018){Jayasinghe}, {Kochanek}, {Stanek},
  {Shappee}, {Holoien}, {Thompson}, {Prieto}, {Dong}, {Pawlak}, {Shields},
  {Pojmanski}, {Otero}, {Britt}, \& {Will}}]{jayasinghe2018a}
{Jayasinghe}, T., {Kochanek}, C.~S., {Stanek}, K.~Z., {et~al.} 2018, \mnras,
  477, 3145, \dodoi{10.1093/mnras/sty838}

\bibitem[{Jones {et~al.}(2001--)Jones, Oliphant, Peterson, {et~al.}}]{scipy}
Jones, E., Oliphant, T., Peterson, P., {et~al.} 2001--, {SciPy}: Open source
  scientific tools for {Python}.
\newblock \url{http://www.scipy.org/}

\bibitem[{{Law} {et~al.}(2016){Law}, {Cherinka}, {Yan}, {Andrews}, {Bershady},
  {Bizyaev}, {Blanc}, {Blanton}, {Bolton}, {Brownstein}, {Bundy}, {Chen},
  {Drory}, {D'Souza}, {Fu}, {Jones}, {Kauffmann}, {MacDonald}, {Masters},
  {Newman}, {Parejko}, {S{\'a}nchez-Gallego}, {S{\'a}nchez}, {Schlegel},
  {Thomas}, {Wake}, {Weijmans}, {Westfall}, \& {Zhang}}]{law2016}
{Law}, D.~R., {Cherinka}, B., {Yan}, R., {et~al.} 2016, \aj, 152, 83,
  \dodoi{10.3847/0004-6256/152/4/83}

\bibitem[{Munich \& Perona(1999)}]{munich1999}
Munich, M., \& Perona, P. 1999, in Proceedings of the Seventh IEEE
  International Conference on Computer Vision, Vol.~1, 108--115 vol.1,
  \dodoi{10.1109/ICCV.1999.791205}

\bibitem[{Oliphant(2007)}]{numpy}
Oliphant, T.~E. 2007, Computing in Science Engineering, 9, 10,
  \dodoi{10.1109/MCSE.2007.58}

\bibitem[{P\'erez \& Granger(2007)}]{ipython}
P\'erez, F., \& Granger, B.~E. 2007, Computing in Science and Engineering, 9,
  21, \dodoi{10.1109/MCSE.2007.53}

\bibitem[{{Petersburg} {et~al.}(2020){Petersburg}, {Ong}, {Zhao}, {Blackman},
  {Brewer}, {Buchhave}, {Cabot}, {Davis}, {Jurgenson}, {Leet}, {McCracken},
  {Sawyer}, {Sharov}, {Tronsgaard}, {Szymkowiak}, \&
  {Fischer}}]{petersburg2020}
{Petersburg}, R.~R., {Ong}, J.~M.~J., {Zhao}, L.~L., {et~al.} 2020, \aj, 159,
  187, \dodoi{10.3847/1538-3881/ab7e31}

\bibitem[{Pickering {et~al.}(2022)Pickering, ojustino, Busko, Conroy, Lim,
  Torres, Nguyen, jehturner, Crawford, Tollerud, Fix, Crawford, Davenport, \&
  Prichard}]{specreduce}
Pickering, T.~E., ojustino, Busko, I., {et~al.} 2022, astropy/specreduce: First
  Official Release, v1.0.0,  Zenodo, \dodoi{10.5281/zenodo.6608788}

\bibitem[{{Samara} {et~al.}(2022){Samara}, {Laperre}, {Kieokaew}, {Temmer},
  {Verbeke}, {Rodriguez}, {Magdaleni{\'c}}, \& {Poedts}}]{samara2022}
{Samara}, E., {Laperre}, B., {Kieokaew}, R., {et~al.} 2022, \apj, 927, 187,
  \dodoi{10.3847/1538-4357/ac4af6}

\bibitem[{Senin(2008)}]{senin2008}
Senin, P. 2008, Information and Computer Science Department University of
  Hawaii at Manoa Honolulu, USA, 855, 40.
\newblock \url{https://csdl.ics.hawaii.edu/techreports/2008/08-04/08-04.pdf}

\bibitem[{{Stoughton} {et~al.}(2002){Stoughton}, {Lupton}, {Bernardi},
  {Blanton}, {Burles}, {Castander}, {Connolly}, {Eisenstein}, {Frieman},
  {Hennessy}, {Hindsley}, {Ivezi{\'c}}, {Kent}, {Kunszt}, {Lee}, {Meiksin},
  {Munn}, {Newberg}, {Nichol}, {Nicinski}, {Pier}, {Richards}, {Richmond},
  {Schlegel}, {Smith}, {Strauss}, {SubbaRao}, {Szalay}, {Thakar}, {Tucker},
  {Vanden Berk}, {Yanny}, {Adelman}, {Anderson}, {Anderson}, {Annis},
  {Bahcall}, {Bakken}, {Bartelmann}, {Bastian}, {Bauer}, {Berman},
  {B{\"o}hringer}, {Boroski}, {Bracker}, {Briegel}, {Briggs}, {Brinkmann},
  {Brunner}, {Carey}, {Carr}, {Chen}, {Christian}, {Colestock}, {Crocker},
  {Csabai}, {Czarapata}, {Dalcanton}, {Davidsen}, {Davis}, {Dehnen},
  {Dodelson}, {Doi}, {Dombeck}, {Donahue}, {Ellman}, {Elms}, {Evans}, {Eyer},
  {Fan}, {Federwitz}, {Friedman}, {Fukugita}, {Gal}, {Gillespie}, {Glazebrook},
  {Gray}, {Grebel}, {Greenawalt}, {Greene}, {Gunn}, {de Haas}, {Haiman},
  {Haldeman}, {Hall}, {Hamabe}, {Hansen}, {Harris}, {Harris}, {Harvanek},
  {Hawley}, {Hayes}, {Heckman}, {Helmi}, {Henden}, {Hogan}, {Hogg}, {Holmgren},
  {Holtzman}, {Huang}, {Hull}, {Ichikawa}, {Ichikawa}, {Johnston}, {Kauffmann},
  {Kim}, {Kimball}, {Kinney}, {Klaene}, {Kleinman}, {Klypin}, {Knapp},
  {Korienek}, {Krolik}, {Kron}, {Krzesi{\'n}ski}, {Lamb}, {Leger},
  {Limmongkol}, {Lindenmeyer}, {Long}, {Loomis}, {Loveday}, {MacKinnon},
  {Mannery}, {Mantsch}, {Margon}, {McGehee}, {McKay}, {McLean}, {Menou},
  {Merelli}, {Mo}, {Monet}, {Nakamura}, {Narayanan}, {Nash}, {Neilsen},
  {Newman}, {Nitta}, {Odenkirchen}, {Okada}, {Okamura}, {Ostriker}, {Owen},
  {Pauls}, {Peoples}, {Peterson}, {Petravick}, {Pope}, {Pordes}, {Postman},
  {Prosapio}, {Quinn}, {Rechenmacher}, {Rivetta}, {Rix}, {Rockosi}, {Rosner},
  {Ruthmansdorfer}, {Sandford}, {Schneider}, {Scranton}, {Sekiguchi}, {Sergey},
  {Sheth}, {Shimasaku}, {Smee}, {Snedden}, {Stebbins}, {Stubbs}, {Szapudi},
  {Szkody}, {Szokoly}, {Tabachnik}, {Tsvetanov}, {Uomoto}, {Vogeley}, {Voges},
  {Waddell}, {Walterbos}, {Wang}, {Watanabe}, {Weinberg}, {White}, {White},
  {Wilhite}, {Wolfe}, {Yasuda}, {York}, {Zehavi}, \& {Zheng}}]{stoughton2002}
{Stoughton}, C., {Lupton}, R.~H., {Bernardi}, M., {et~al.} 2002, \aj, 123, 485,
  \dodoi{10.1086/324741}

\bibitem[{{Tody}(1986)}]{tody1986}
{Tody}, D. 1986, in Society of Photo-Optical Instrumentation Engineers (SPIE)
  Conference Series, Vol. 627, Instrumentation in astronomy VI, ed. D.~L.
  {Crawford}, 733, \dodoi{10.1117/12.968154}

\bibitem[{{Tran} {et~al.}(2019){Tran}, {Tuttle}, {Mckay}, {Kadlec}, \&
  {Sayres}}]{tran2019}
{Tran}, D., {Tuttle}, S., {Mckay}, M., {Kadlec}, K., \& {Sayres}, C. 2019, in
  American Astronomical Society Meeting Abstracts, Vol. 233, American
  Astronomical Society Meeting Abstracts \#233, 146.07

\bibitem[{{Valdes}(1996)}]{valdes1996}
{Valdes}, F.~G. 1996, in Astronomical Society of the Pacific Conference Series,
  Vol. 101, Astronomical Data Analysis Software and Systems V, ed. G.~H.
  {Jacoby} \& J.~{Barnes}, 33

\bibitem[{Velichko \& Zagoruyko(1970)}]{velichko1970}
Velichko, V., \& Zagoruyko, N. 1970, International Journal of Man-Machine
  Studies, 2, 223, \dodoi{https://doi.org/10.1016/S0020-7373(70)80008-6}

\bibitem[{{W}es {M}c{K}inney(2010)}]{pandas}
{W}es {M}c{K}inney. 2010, in {P}roceedings of the 9th {P}ython in {S}cience
  {C}onference, ed. {S}t\'efan van~der {W}alt \& {J}arrod {M}illman, 56 -- 61,
  \dodoi{10.25080/Majora-92bf1922-00a}

\bibitem[{{York} {et~al.}(2000){York}, {Adelman}, {Anderson}, {Anderson},
  {Annis}, {Bahcall}, {Bakken}, {Barkhouser}, {Bastian}, {Berman}, {Boroski},
  {Bracker}, {Briegel}, {Briggs}, {Brinkmann}, {Brunner}, {Burles}, {Carey},
  {Carr}, {Castander}, {Chen}, {Colestock}, {Connolly}, {Crocker}, {Csabai},
  {Czarapata}, {Davis}, {Doi}, {Dombeck}, {Eisenstein}, {Ellman}, {Elms},
  {Evans}, {Fan}, {Federwitz}, {Fiscelli}, {Friedman}, {Frieman}, {Fukugita},
  {Gillespie}, {Gunn}, {Gurbani}, {de Haas}, {Haldeman}, {Harris}, {Hayes},
  {Heckman}, {Hennessy}, {Hindsley}, {Holm}, {Holmgren}, {Huang}, {Hull},
  {Husby}, {Ichikawa}, {Ichikawa}, {Ivezi{\'c}}, {Kent}, {Kim}, {Kinney},
  {Klaene}, {Kleinman}, {Kleinman}, {Knapp}, {Korienek}, {Kron}, {Kunszt},
  {Lamb}, {Lee}, {Leger}, {Limmongkol}, {Lindenmeyer}, {Long}, {Loomis},
  {Loveday}, {Lucinio}, {Lupton}, {MacKinnon}, {Mannery}, {Mantsch}, {Margon},
  {McGehee}, {McKay}, {Meiksin}, {Merelli}, {Monet}, {Munn}, {Narayanan},
  {Nash}, {Neilsen}, {Neswold}, {Newberg}, {Nichol}, {Nicinski}, {Nonino},
  {Okada}, {Okamura}, {Ostriker}, {Owen}, {Pauls}, {Peoples}, {Peterson},
  {Petravick}, {Pier}, {Pope}, {Pordes}, {Prosapio}, {Rechenmacher}, {Quinn},
  {Richards}, {Richmond}, {Rivetta}, {Rockosi}, {Ruthmansdorfer}, {Sandford},
  {Schlegel}, {Schneider}, {Sekiguchi}, {Sergey}, {Shimasaku}, {Siegmund},
  {Smee}, {Smith}, {Snedden}, {Stone}, {Stoughton}, {Strauss}, {Stubbs},
  {SubbaRao}, {Szalay}, {Szapudi}, {Szokoly}, {Thakar}, {Tremonti}, {Tucker},
  {Uomoto}, {Vanden Berk}, {Vogeley}, {Waddell}, {Wang}, {Watanabe},
  {Weinberg}, {Yanny}, \& {Yasuda}}]{york2000}
{York}, D.~G., {Adelman}, J., {Anderson}, Jr., J.~E., {et~al.} 2000, \aj, 120,
  1579, \dodoi{10.1086/301513}

\bibitem[{Zou {et~al.}(2019)Zou, Zhu, Shen, He, Su, Fan, Lu, Zhang, \&
  Chen}]{zou2019}
Zou, C., Zhu, H., Shen, J., {et~al.} 2019, Anal. Methods, 11, 4481,
  \dodoi{10.1039/C9AY01139K}

\end{thebibliography}

\end{document}